\documentclass{aa}
\usepackage{epsfig}
\usepackage{rotating}
\usepackage{graphicx}
\newcommand{\rosn}{{\sl R\"ontgen Satellite}}
\newcommand{\ros}{{\em ROSAT}}
\newcommand{\chan}{{\em Chandra}}
\newcommand{\xmm}{{\em XMM-Newton}}
\newcommand{\vltn}{{\em Very Large Telescope}}
\newcommand{\vlt}{{\em VLT}}
\newcommand{\lbtn}{{\em Large Binocular Telescope}}
\newcommand{\lbt}{{\em LBT}}
\newcommand{\hstn}{{\em  Hubble  Space Telescope}}
\newcommand{\hst}{{\em HST}}

\newcommand{\fors}{{\em FORS2}}
\newcommand{\forsn}{{\em FOcal Reducer/low dispersion Spectrograph}}

\def \oneight{RX\, J1856.5$-$3754}
\def \zeroseven{RX\, J0720.4$-$3125}

\def \onethree{RX\, J1308.6+2127}

\def \oneseven{1RXS\, J214303.7+065419}
\def \rbs{RBS\, 1774}

\begin{document}

\title{VLT/FORS2 observations of the optical counterpart of the isolated neutron star RBS\, 1774\thanks{Based on observations collected at ESO, Paranal, under Programme 383.D-0485(A)}}

\author{R. P. Mignani\inst{1,2}
\and
S. Zane \inst{1}
\and
R. Turolla\inst{3,1}
\and
F. Haberl\inst{4}
\and 
M. Cropper\inst{1}
\and
C. Motch\inst{5}
\and
A. Treves\inst{6}
\and
L. Zampieri\inst{7}
}

   \institute{Mullard Space Science Laboratory, University College London, Holmbury St. Mary, Dorking, Surrey, RH5 6NT, UK
   \and{Institute of Astronomy, University of Zielona G\'ora, Lubuska 2, 65-265 Zielona G\'ora, Poland}
\and
Department of Physics, University of Padua, via Marzolo 8,  Padua, 35131, Italy 
\and
CNRS, Universit\'e de Strasbourg, Observatoire Astronomique, 11 rue de l'Universit\'e, 67000 Strasbourg, France 
\and 
Max Planck Institut f\"ur Extraterrestrische Physik, Giessenbachstrasse, D85748, Garching, Germany
\and
Dipartimento di Fisica e Matematica, Universit\'a dell'Insubria, via Valleggio 11, 22100, Como, Italy
\and
INAF, Osservatorio Astronomico di Padova, Vicolo dell'Osservatorio 5, I-35122, Padova, Italy
}

\titlerunning{\vlt\ observations of RBS\, 1774}

\authorrunning{Mignani et al.}
\offprints{R. P. Mignani; rm2@mssl.ucl.ac.uk}

\date{Received ...; accepted ...}

\abstract{X-ray observations performed with the \rosn\ (\ros) led to the discovery of a group (seven to date) of X-ray dim  and radio-silent middle-aged isolated neutron stars (a.k.a.  XDINSs), which are characterised  by pure  blackbody spectra  ($kT\approx 40-100$ eV), long X-ray pulsations ($P=3-12$ s),  and appear to be endowed  with  relatively  high magnetic  fields, ($B\approx  10^{13}$--$10^{14}$~G). Optical observations of XDINSs are important, together with the X-ray ones, to study the cooling of the neutron star surface and to investigate the relation betwen XDINSs and other isolated neutron star classes. RBS\, 1774
is one of the few XDINSs with a candidate optical counterpart, which we discovered with the \vltn\ (\vlt).
}
{We aim at  constraining the optical spectrum of  RBS\, 1774, for which only two B-band flux measurements
are available,  and to determine whether its optical emission is either of thermal or of non-thermal origin.}   
{We performed deep observations of RBS\, 1774 in the R band with the
\vlt\ to  disentangle a non-thermal power-law spectrum from a Rayleigh-Jeans, whose contributions are expected to be very much different in the red part of the spectrum.}
{We did not detect the RBS\, 1774 candidate counterpart
down to a $3 \sigma$ limiting magnitude of $R\sim 27$. The constraint on its colour, $(B-R)\la0.6$, rules out that it is a background object, positionally coincident with the X-ray source.  Our R-band  upper limit is consistent with the extrapolation of the B-band flux (assuming a $3 \sigma$ uncertainty) for a set of power-laws $F_{\nu} \propto \nu^{-\alpha}$ with spectral indeces $\alpha \le 0.07$. If the optical spectrum of \rbs\ were non-thermal, its power-law slope would be very much unlike those of all isolated neutron stars  with non-thermal optical emission, suggesting that it is most likely thermal. For instance, a Rayleigh-Jeans  with temperature $T_O = 11$~eV, for an optically emitting radius $r_O=15$~km and a source distance $d=150$~pc,  would be consistent with the optical measurements. The implied low distance is compatible with the 0.04 X-ray pulsed fraction  if either the star spin axis is nearly aligned with the magnetic axis  or  with  the line  of sight, or it is slightly misaligned with respect to both the magnetic axis and the line of sight by $5-10^{\circ}$.  }
{New observations, both from the ground and from the \hstn\ (\hst), are important to  characterise the optical/near-ultraviolet (UV) spectrum of \rbs, to better constrain the values of $r_O$, $d$, and $T_O$ and measure the source proper motion from which indirect constraints on the source distance can be inferred.}

 
   \maketitle

\section{Introduction}

One of the most intriguing results  of the all sky survey performed by the  \rosn\  (\ros) has  been  the  discovery  of seven X-ray dim and  radio-silent, middle-aged  isolated neutron stars (a.k.a.  XDINSs, see Haberl 2007 and  Turolla 2009, for  the most recent reviews).  XDINSs  stand apart with respect to most X-ray emitting  INSs due to their pure blackbody X-ray spectra  ($kT\approx 40-100$  eV), produced  from the  cooling  of the neutron star surface. 
Shallow  X-ray pulsations  ($P=3-12$ s), likely  from large and  hot polar  caps,  are observed  for  all but  one  of them.   The measurement of  the period derivative  $\dot {P}$ in some  XDINSs (see Kaplan \& van Kerkwijk  2009 and references therein) yielded, assuming
magneto-dipolar spin-down,  ages of  $\sim 1-2$ Myrs  and rotational energy losses $\dot {E} \sim 2-5 \times 10^{30}$ ergs s$^{-1}$, too small to  power detectable  magnetospheric  emission. Interestingly,  XDINSs appear to  be endowed with relatively high  magnetic fields, $B\approx 10^{13}$--$10^{14}$~G, as inferred from  spin-down measurements and the detection  of broad  spectral features  which could  be  attributed to proton cyclotron scattering and/or electron transitions in H/He-like atoms.
Besides  the X-rays,  XDINSs are only  detected in  the optical. In radio, stringent upper limits have been obtained  at  820 MHz
(Kondatriev  et al. 2009),  while the detections  of \onethree\ and  \oneseven\  at 111 MHz (e.g. Malofeev et al. 2007) is yet unconfirmed. XDINSs feature  a $\ge 5$ optical excess with respect to the  extrapolation of the  X-ray spectrum (see, e.g.  Mignani 2011) and,  at  least  in   the  two  best-studied  cases  (\zeroseven\  and \oneight), their optical fluxes  closely follow a Rayleigh-Jeans (R-J) distribution, with a possible additional power-law (PL) component for the former (Kaplan et al. 2003).  Whether  the XDINSs  optical emission is  produced from regions of  the star  surface which are  larger and cooler  than those producing the X-rays (e.g.  Pons et al.  2002) is debated.

Recently, we  observed \oneseven\ (a.k.a. RBS\, 1774),  the last of the \ros\ discovered XDINS (Zampieri et al. 2001) and possibly that with
the highest inferred magnetic  field ($\approx 10^{14}$~G; Zane et al. 2005)\footnote{This value has been inferred from the observations of an absorption feature in the X-ray spectrum of the source.  However, the recent measurement of the pulsar spin-down (Kaplan \& van Kerkwijk 2009) implies a magnetic field of $\sim 2 \times 10^{13}$ G},  with   the  \vltn\  (\vlt)  and  we   identified  its  optical counterpart (B=27.4$\pm$ 0.2; Zane et al.   2008).  
Interestingly,  the neutron star  flux was  found to  be a factor  $\approx 35$  above the optical extrapolation  of the \xmm\  X-ray spectrum ($kT\sim  104$ eV; $N_H\sim 3.6\times 10^{20}$ ${\rm cm}^{-2}$;  Zane et al. 2005), and a slightly  larger optical  excess was  
measured  with the \lbtn\ (\lbt)  by Schwope et  al.  (2009).   
So far, nothing is known  about the optical/IR spectrum of RBS\, 1774, for which only  the two B-band flux measurements of Zane et al. (2008) and Schwope et al. (2009)  are  available.  Prior to our  detection  of  the  optical  counterpart, the  neutron  star  was
observed both  at optical and infrared (IR) wavelengths (Lo  Curto et al. 2007; Rea et al. 2007; Posselt et al. 2009) but no counterpart was found.

In  order  to constrain  the  optical  spectrum  of  RBS\, 1774,  we  performed  new observations  with the  \vlt\  in the  R  band
to disentangle  the slope of  a  flat PL continuum from  that of  a R-J distribution and,  thus, to determine whether the  optical emission is non-thermal or thermal, i.e. associated either with 
 the magnetosphere or 
the neutron  star surface.

This paper  is organised as follows: observations,  data reduction and
analysis are  described in  Sect. 2, while  results are  presented and
discussed in Sect. 3 and. Conclusions follow.

\section{Observations and data analysis}

\subsection{Observation description}

We obtained optical images of the RBS\, 1774 field with the \vlt\ Antu
telescope at the ESO Paranal  observatory on September 19 and 20, 2009.
Observations were  performed in service  mode with the \forsn\  (\fors), a
multi-mode camera for  imaging and long-slit/multi-object spectroscopy
(Appenzeller  et  al. 1998),  using  the  default  $R Special$  filter
($\lambda=6550$ \AA;  $\Delta \lambda=1650$\AA).  In  order to achieve
the  highest sensitivity  at longer  wavelengths, \fors\  was equipped
with its red-sensitive MIT detector, a mosaic of two 2k$\times$4k CCDs
optimised  for wavelengths  longer  than 6000  \AA.   In its  standard
resolution  mode,   the  detector  has  a  pixel   size  of  0\farcs25
(2$\times$2 binning) which  corresponds to a projected field--of--view
of 8$\farcm3  \times 8\farcm3$  over the CCD  mosaic. However,  due to
vignetting, the effective sky coverage of the two detectors is smaller
than the projected detector field--of--view,  and it is larger for the
upper CCD chip.   To include a larger number of  reference stars for a
precise image astrometry, as well as to increase the signal--to--noise
ratio per pixel, we  performed the observations in standard resolution
mode.  We positioned  RBS\, 1774 in the upper CCD  chip to exploit its
larger  effective  sky  coverage  ($7\arcmin \times  4\arcmin$).   The
instrument was operated in its standard low gain, fast read-out mode.

To allow for  cosmic ray removal, a sequence of 20  exposures of 775 s
each were obtained in the two  nights, for a total integration time of 9300 s and 
6200  s for the first and second night, respectively. Exposures  were taken  in  dark time  and under  photometric
conditions,    as   recorded    by   the    ESO    ambient   condition
monitor\footnote{http://archive.eso.org/asm/ambient-server}  and close
to the  zenith, with an airmass  mostly below 1.3.   The image quality
(IQ)  was not  constant during  our observations.   The  seeing values
measured by  the Paranal differential image motion  monitor (DIMM) are
relative  to the  zenith  and not  to  the pointing  direction of  the
telescope  and  are  subject  to  ground-layer  turbulence  (see  also
Martinez et al.  2010).  Thus,  they are not necessarily indicative of
the actual IQ. For this reason, we computed it from the measured point
spread function (PSF), derived by  fitting the full width half maximum
(FWHM)  of  a  number  of  well-suited  field  stars  using  the  {\em
Sextractor} tool  (Bertin \& Arnouts 1996), following  the recipe used
by        the         \fors\        data        quality        control
procedures\footnote{www.eso.org/observing/dfo/quality/FORS2/qc/qc1.html}. The
average IQ in the first night was $\sim 0\farcs6$, while in the second
night it was $\sim 0\farcs9$.  Bias, twilight flat--fields frames, and
images of standard star fields (Landolt 1992) were obtained as part of
the \fors\ science calibration plan.

\subsection{Data analysis}

Data   were   reduced   through   the  ESO   \fors\   data   reduction
pipeline\footnote{www.eso.org/observing/dfo/quality/FORS2/pipeline}
for  bias   subtraction,  and  flat--field   correction.   Photometric
calibration  was  applied using  the  extinction-corrected night  zero
points  computed by  the  \fors\ pipeline  and  available through  the
instrument  data quality  control database.   We converted  the \fors\
zero points,  computed in units of  electrons/s, to units  of ADU/s by
applying the corresponding electrons--to--ADU conversion factors.
We finally  co-added the reduced  science images using the  {\em IRAF}
task {\tt  drizzle} applying a $3  \sigma$ filter on  the single pixel
average  to filter  out residual  hot and  cold pixels.   In  order to
achieve a better signal--to--noise, we  decided to use only the images
taken on the first night, which have the best IQ in our sample and provide the longest integration time (9300 s). 
Of course, we verified that adding the images taken on the second night does not improve the signal--to--noise.

To locate the RBS\, 1774 candidate counterpart in our \fors\ image, we
used as a reference its  position measured in our original {\em FORS1}
observation. Since the  {\em FORS1} detector was also  a mosaic of two
2k$\times$4k CCDs  (but blue-optimised), with the same  pixel size and
field--of--view  as  \fors\  at  standard resolution,  we  decided  to
register the counterpart position on our co-added \fors\ image through
a  relative astrometry  procedure.  Whenever  possible,  this approach
yields a  position registration in the detector  reference frame which
is  more accurate  than that  achievable through  absolute astrometry,
since it  is not  affected by the  intrinsic absolute accuracy  of the
used reference catalogue. Moreover,  most of the reference stars from
e.g. the GSC-2 (Lasker et al.  2008) are saturated in our images.  For
our  relative astrometry,  we used  as  a reference  the positions  of
several point-like  objects detected in the  field--of--view with {\em
Sextractor}, after filtering out objects that are either saturated, or
too  faint,  or  too close  to  the  CCD  edges.  The  relative  frame
registration  was performed  using  standard tools  available in  {\em
IRAF} and turned out to be  accurate to better than 0.05 pixels ($\sim
0\farcs0125$) in  both the $x$  and $y$ directions, which  are aligned
with right ascension and declination to better than $0.1^{\circ}$.  To
this, we  added the uncertainty on the  candidate counterpart centroid
on  the  {\em   FORS1}  image  which  is  $\sim   0.1$  pixels  ($\sim
0\farcs025$). Since  the {\em FORS1}  position refers to July  11, 2007
(MJD=54292),  we   have  to  account  for   an  additional  positional
uncertainty due to the unknown  neutron star proper motion to the date
of our \fors\ observations  (MJD=55093).  This corresponds to a yearly
displacement of $\sim  0.2\arcsec v_{100}/d_{100}$, where $v_{100}$
and $d_{100}$ are the unknown neutron star transverse velocity and distance\footnote{Posselt et al. (2009) quote a fiducial lower limit of $300$ pc, estimated from models of the $N_H$ distribution.} in
units  of 100  km  s$^{-1}$  and 100  pc,  respectively.  For, e.g. an
average  neutron star  velocity of  $\sim 400$  km s$^{-1}$  (Hobbs et
al. 2005) and a distance of 300 pc, this would translate
into  an additional  positional  uncertainty of  $\sim 0\farcs6$,  which obviously dominates over the uncertainty on our relative astrometry.

\begin{figure}
\includegraphics[height=8.5cm]{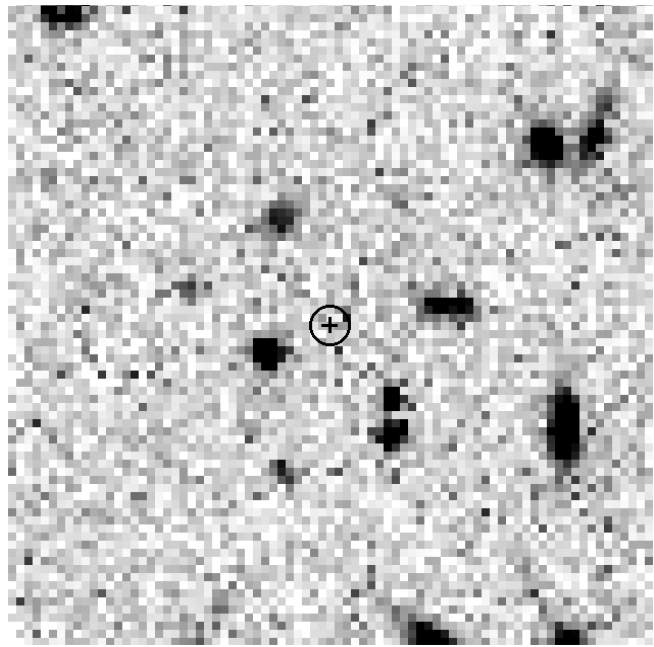}
  \caption{\vlt/\fors\  $20\arcsec \times 20\arcsec$ co-added R-band image (9300 s integration time) of  the RBS\, 1774 field (North to the top, east to the left). The computed position of the candidate counterpart is marked by the cross. For a better visualisation, the cross arms  (0\farcs25) are equal to 20 times the $1 \sigma$ uncertainty of our relative astrometry. The circle ($\sim 0\farcs6$ radius) corresponds to  the position uncertainty due to the unknown neutron star proper motion for, e.g. a transverse velocity of 400 km s$^{-1}$ and a distance of 300 pc. }
\end{figure}

\section{Results}

\subsection{Photometry}

The  computed  position  of  the  RBS\,  1774  candidate  counterpart,
overlaid on the co-added \fors\ R-band  image, is shown in Fig. 1.  As
it is apparent,  no source can be visually  identified at the expected
neutron  star   position.  We  ran  an  automatic   search  with  {\em
Sextractor}  but we did  not detect  any object  above the  $3 \sigma$
detection threshold.   We applied an image smoothing  using a Gaussian
function over $3 \times 3$ pixel cells but we could not detect any new
feature which could  be unambiguously associated to a  real object and
not to an enhanced background fluctuation.  Thus, we conclude that the
optical counterpart  of RBS\,  1774 is undetected  in our  images.  We
note  that  the  0\farcs6  positional  uncertainty  shown  in  Fig.  1
corresponds to  the assumed  values of the  neutron star  distance (300 pc) and
transverse velocity (400 km s$^{-1}$) and, thus, it is is only indicative.  However, our
conclusions are not affected by the actual position uncertainty of the
neutron  star since  no  new object  is  detected within  a radius  of
5\arcsec\ from the nominal candidate counterpart position with respect
to  those already  detected (see  Fig.  1 of  Zane et  al. 2008).   We
determined the detection limit of  our \fors\ image from the number of
counts corresponding  to a  $3 \sigma$ detection  in an aperture  of 2
pixel radius  (0\farcs5), derived from  the standard deviation  of the
background sampled around the expected neutron star position.  We then
corrected the number of counts  for the airmass, using the atmospheric
extinction  coefficients measured for  \fors\ and available through  the
instrument  data quality  control database,
and we
applied  the aperture  correction. The  latter was  computed  from the
measured growth curve of a number of relatively bright but unsaturated
stars in  the field, with no  adjacent star closer  than 5\arcsec, and
whose photometry is not  affected by strong background gradients.  The
final $3 \sigma$  limit of $R \sim 27$  is about 1.3 magnitudes
fainter than the $r'$ upper limit reported by Rea et al. (2007).

The non-detection of the Zane et al. (2008) candidate in our new \vlt\
R-band images might cast some  doubts as whether this object is indeed
the  optical  counterpart  of  RBS\,  1774, although  the  low  chance
coincidence probability with the  \chan\ position ($2 \times 10^{-3}$;
Zane et  al. 2008) already  made the association quite  robust.  Given
the  flux  of  the  candidate  counterpart  ($B=27.4  \pm  0.2$),  its
non-detection in the R  band corresponds to an observed $(B-R)\la0.6$.
The observed upper limit on the colour,
together with the brightness of the candidate counterpart, can be used
to rule out that this  is a background object, positionally coincident
with the neutron  star. Since RBS\, 1774 coordinates  point at a quite
high galactic latitude ($l=62.6556^{\circ}$, $b=-33.1392^{\circ}$), as
shown  by the  many galaxies  identified in  the field,  the candidate
counterpart  might be,  in principle,  an AGN.   Indeed, most  AGNs at
$z\la2$ have $0\la  (B-R) \la 3$, which is  compatible with the colour
of   RBS\,   1774   candidate    counterpart.   The   fact   that   no
transient/persistent  radio source  was  detected at  its position  in
recent, deep 840 MHz observations  (Kondriatev et al. 2009) would only
rule  out  a  radio-loud  AGN.    However,  an  AGN  not  detected  in
X-rays\footnote{\chan\  observations of  RBS\, 1774  rule out  a blend
with a background X-ray source.} would be either very much absorbed or
at a much  higher redshift, which would be  hardly compatible with the
constraint  on its  $(B-R)$  and with  its  detection in  the B  band,
respectively.  On the other  hand, a $(B-R)\la0.6$ would be compatible
with a  main sequence star  of spectral type  earlier than F  but this
should be  well outside  the Galaxy to  reproduce the  observed B-band
flux.  Thus, both  the counterpart colour and its  brightness are only
compatible with a neutron star.

\begin{table*}
\caption{Best-fit parameters to the X-ray \xmm\ spectrum computed for different models.}

\begin{tabular}{lccccc}
\hline\hline\noalign{\smallskip}
\multicolumn{1}{l}{Model} &
\multicolumn{1}{c}{N$_{\rm H}$} &
\multicolumn{1}{c}{kT} &
\multicolumn{1}{c}{E$_{\rm line}$} &
\multicolumn{1}{c}{$\sigma_{line}$} &
\multicolumn{1}{c}{$\chi_r2$(dof)}\\
\multicolumn{1}{l}{} &
\multicolumn{1}{c}{($10^{20}$ cm$^{-2}$)} &
\multicolumn{1}{c}{(eV)} &
\multicolumn{1}{c}{(eV)} &
\multicolumn{1}{c}{(eV)} &
\multicolumn{1}{c}{} \\

\noalign{\smallskip}\hline\noalign{\smallskip}
(1) bbody        &    2.7$\pm$0.2       & 102.0$\pm$1.2 &     $-$        &    $-$          &   1.51 (133)    \\
(2) bbody+gauss  &    2.6$\pm$0.2       & 104.2$\pm$2.1 &  728$\pm$16    &  31$\pm$31      &   1.20 (130)    \\
(3) bbody*gauss  &    2.6$\pm$0.2       & 104.1$\pm$2.1 &  733$\pm$17    &  31$\pm$30      &   1.20 (130)    \\

\noalign{\smallskip}\hline\noalign{\smallskip}
\end{tabular}
\end{table*}
\subsection{Multi-band spectrum}

We compared  our $R$-band magnitude  upper limit with  the
best-fit  model to  the  \xmm\ spectrum  of  RBS\, 1774  and with  the
available multi-band  photometry (Rea et  al. 2007; Zane et  al. 2008;
Schwope  et  al.   2009).   We computed  the  interstellar  extinction
correction  from the hydrogen  column density  $N_H$ derived  from the
X-ray spectral fit.
To  this  aim,  we  fitted  the  \xmm\  spectrum  anew  using  updated
calibrations and  response files. 
We re-extracted the EPIC-pn spectrum from the original May 2004 observation 
using version 10.0.0 of the \xmm\ Science Analysis System (SAS),
 a circular region of  
$20\arcsec$ radius, and selecting only single-pixel events. Since 
combined fits with pn and MOS spectra are dominated by the 
higher statistical quality of the pn spectrum, we concentrated 
on the pn spectrum only.
For comparison with the results obtained by Schwope et al. (2009) and 
Cropper et al. (2007) we fitted the spectrum with either  a black body (BB)
model, a BB plus an (additive) Gaussian line, and a BB plus a 
(multiplicative) absorption edge. 
In each case, interstellar absorption was included by using the {\tt XSPEC} (version 12.6.0k) 
model {\it phabs} with abundances from Wilms et al. (2000). The best fit parameters 
are summarised in Table 1. 
As in previous analyses, inclusion of an absorption feature improves the fit over a simple BB model, although the current energy resolution does not allow to discriminate a Gaussian line in absorption from an edge.
The slight differences between the model parameters reported here and those 
given by Schwope et al. (2009) and Cropper et al. (2007) are largely due to the 
different revisions of the soft spectral response calibration of the EPIC-pn camera
and possibly different elemental abundances used for the absorption model.
 The  new spectral fit with an absorbed BB plus an (additive) Gaussian line yields a best-fit temperature of $kT =
104.2 \pm  2.1$ eV, consistent with  that obtained by  Zane et al.
(2005) by fitting the same \xmm\  data but with older calibrations and
response files.  However,  we note that our new  spectral fit yields a
best-fit $N_H$  value of $(2.6  \pm 0.2) \times  10^{20}$ cm$^{-2}$
which  is somewhat lower  than that  obtained by  Zane et  al. (2005),
$N_H= 3.6 \times 10^{20}$ cm$^{-2}$, and which was used both in Rea et
al.  (2007)  and in Zane et  al.  (2008) to  compute the interstellar
extinction.
This resulted
in slightly  larger extinction-corrected  fluxes, where the  effect is
$\la  0.08$ magnitudes,  i.e. well  below the  uncertainties  on their
absolute  photometry.  Nonetheless, in  the following  we assume  as a
reference  $N_H= 2.6 \times 10^{20}$ cm$^{-2}$,  obtained from our
updated  spectral fits  to the  \xmm\ spectrum.   From this  value, we
derived an interstellar reddening $E(B-V)=0.046$ using the relation of
Predehl \& Schmitt (1995).\footnote{This 
relation is affected by 
uncertainties for close objects, due to 
the problems of modelling the interstellar medium (ISM) at small distance from the Sun where
microstructures weight more. We checked that, when using the relations of 
Bohlin et al. (1978) and of Paresce (1984)
extinction corrections are consistent within 0.04 magnitudes, 
well below 
the
pure statistical error on the source count rate.} and, from this, we computed the interstellar
extinction in the different bands using the extinction coefficients of
Fitzpatrick  (1999).   We  then  corrected  the  available  multi-band
photometry accordingly. 

The new extinction-corrected multi-band fluxes from Rea et al. (2007),
Zane et  al. (2008), and Schwope et  al.  (2009) are shown  in Fig. 2,
together with the best-fit \xmm\ spectrum and its extrapolation in the
optical domain.   To these points, we added  flux measurements obtained
in the near-ultraviolet (UV).  The  \rbs\ field has been observed with
the  {\em GALEX}  satellite  (Martin et  al.   2005) in  both the  NUV
($\lambda=2771$ \AA;  $\Delta \lambda\sim 530$ \AA)  and FUV passbands
($\lambda=1528$ \AA; $\Delta \lambda\sim  220$ \AA).  We retrieved the
fully  reduced  and  calibrated  imaging  data from  the  {\em  GALEX}
archive\footnote{www.galex.stsci.edu} for inspection  but we could not
detect the source down to $3  \sigma$ upper limits of 2.76 $\mu$Jy and
5.77 $\mu$Jy in the NUV and FUV passbands, respectively.  We corrected
these fluxes for the interstellar  extinction using as a reference the
$E(B-V)=0.046$  derived  above  and  the  extinction  coefficients  of
Fitzpatrick (1999) for  the NUV passband and of  Seaton (1979) for FUV
one.  The  field has  been also observed  with the \xmm\  {\em Optical
Minotor} (Mason et al.  2001) in the UWW1 ($\lambda=2675$ \AA; $\Delta
\lambda\sim 577$  \AA), UWM2 ($\lambda=2205$  \AA; $\Delta \lambda\sim
350$ \AA), and UVW2 ($\lambda=1894$ \AA; $\Delta \lambda\sim 330$ \AA)
filters but the source was not detected down to $3\sigma$ limits which
are not deeper than the  {\em GALEX}/NUV one.  At shorter wavelengths,
the source was not detected by  {\em EUVE} during its all-sky scan and
was not targeted by pointed observations (Bowyer et al.  1996). As seen
from  Fig. 2,  the  {\em GALEX}  points  lie at  least  two orders  of
magnitude above  the X-ray spectrum extrapolation. Thus,  they can not
be used to constrain the shape of the optical spectrum.

\begin{figure*}
\begin{center}
\includegraphics[height=8.5cm]{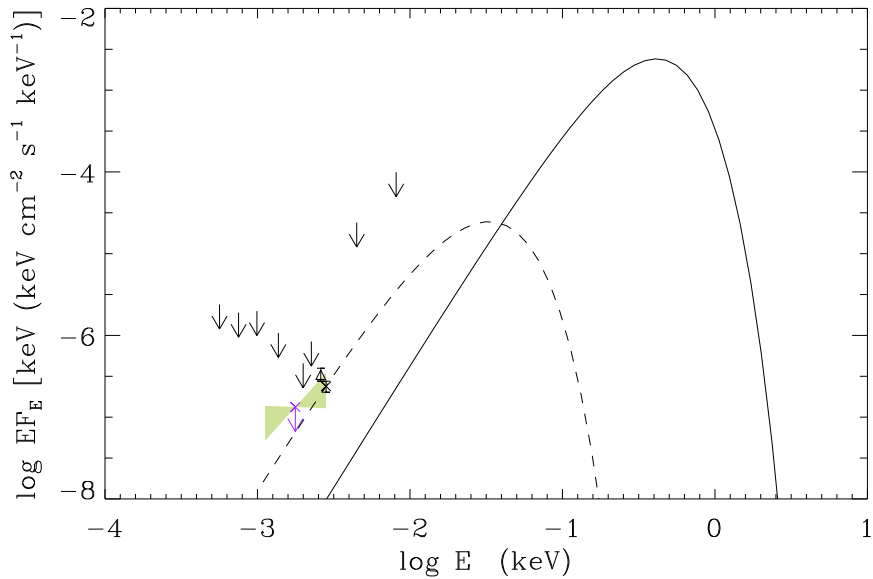}
  \caption{Optical/X-ray  spectral energy  distribution of  RBS\, 1774. 
The solid  line represents the  updated best-fit model spectrum ($kT\sim 104$  eV; $N_H\sim 2.6\times 10^{20}$  ${\rm cm}^{-2}$) to the \xmm\ data. The arrows represent  the 5$\sigma$ optical/IR flux  upper limits reported
by Rea  et al.   (2007) and the  $3 \sigma$  near/far-ultraviolet flux
upper  limits  derived from  archival  {\em  GALEX} observations  (see
text).  The \vlt\ B-band flux  of the RBS\, 1774 candidate counterpart
is marked by  the cross, while our \vlt\ R-band  upper limit is marked
by the cross and the  arrow. The {\em LBT} B-band measurement (Schwope
et al.   2009) is marked by  the triangle. The  shaded areas represent
the region described by the  two PLs with  slopes
$\alpha=-2.07$  and  0.07 which pass  $\pm 3 \sigma$  above/below the B-band flux and 
are consistent with  our R-band  upper  limit   (see text for details).   Optical points  have  been
corrected for the $N_H$ obtained  from the updated fit to the \xmm\
spectrum (see text).   The dashed line represents
the blackbody at $T_O =11$~ eV ($r_O=15$~ km, $d=$150 pc), normalized to our \vlt\ B band point  and compatible with the   R-band upper limit. }
\end{center}
\end{figure*}

\subsection{The optical excess}

We note  that the  extrapolation of the  new best-fit spectrum  in the
optical domain lies a factor  1.35 above 
the extrapolation derived 
 in Zane et al. (2008).
Thus, as the  result of the new  fit to the \xmm\ spectrum
the optical  excess of  the  \vlt\ B-band  flux  
 is $\sim
24.4 \pm 3.7$  ($1 \sigma$ confidence level).  This  is  lower
than the  value of $35  \pm 7$ ($1  \sigma$) published in Zane  et al.
(2008)
although still marginally consistent within $1 \sigma$ errors.   
As shown in Sect. 3.2, the difference in the $N_{\rm H}$ obtained from our spectral fit and that of Zane et al. (2005) only corresponds to a difference of $\la 0.08$ magnitudes in the dereddened B-band flux. Thus, it has only a negligible effect in the lower optical excess that we derive, which is mainly due to the difference in the extrapolation of the best-fit \xmm\  spectrum, resulting  from  the  use  of  new  calibrations  and response  files. We also note that the value of the optical excess is insensitive to which of the two best-fit spectral models, (2) and (3) in Table\, 1, is actually assumed, since both the spectral parameters of the underlying blackbody and the inferred $N_{\rm H}$ are virtually identical. 
We note  that
Schwope et al.  (2009), using  \lbt\ data, obtained a
slightly brighter B-band magnitude of $26.96 \pm 0.20$ with respect to
our  value  of  $B=27.4 \pm  0.2$  (Zane  et  al.  2008),  which  they
interpreted as  hint of a  possible flux variability from  the source.
Taking the magnitude difference at  face value, this would  imply a
difference of  $\approx 50\%$  in the 
optical
excess. However, we note that the magnitude difference between the two
measurements,  $0.44  \pm  0.28$ ($1 \sigma$ error),  is not  statistically  compelling.
Moreover,  the  two  measurements   have  been  taken  with  different
telescope/detectors  and with non-identical  filters, which  brings in an
additional  uncertainty due  to the  colour-term correction,  and have
been calibrated using different  sets of standard stars, using default
atmospheric extinction corrections.  While Zane et al.  (2008) applied
the  average  B-band atmospheric  extinction  coefficient provided  by
ESO\footnote{www.eso.org/observing/dfo/quality/FORS1/qc/qc1.html}
computed from  observations performed at the  Paranal Observatory over
the    April   2007-September    2007   semester    with    the   same
telescope/instrument set-up used in their own observations, Schwope et
al.  (2009) applied the B-band extinction coefficients measured at the
\lbt\  site  a few  months  apart,  during  the commissioning  of  the
instrument.   Since the  measured atmospheric  extinction coefficients
can display a night-to-night scatter up to $\sim$ 0.1 magnitudes, this
would introduce, on both measurements, an additional uncertainty of up
to  $\sim 0.1$  magnitudes on  the absolute  photometry, for  a target
observed at  the zenith,  which is obviously  larger if the  target is
observed at larger zenith angles. Accounting for all the above sources
of  uncertainties,   the  difference  between  the   two  B-band  flux
measurements  are significantly  smoothed out,  and we  regard  them as
consistent  with in the errors.   Thus, in  the following  we use  as a
reference a value of the optical  excess of $\sim 24.4 \pm  3.7$, corresponding to
our \vlt\ B-band measurement.

\section{Discussion}

\subsection{Non-thermal emission}

Our  new R-band  magnitude upper  limit corresponds  to  an unabsorbed spectral  flux  $<1.33  \times  10^{-7}$  keV~(keV~cm$^{-2}$  s$^{-1}$ keV$^{-1}$), at a central energy of  1.77 eV.  As it is seen from Fig. 2,  the R-band upper limit is consistent with a set of  PLs $F_{\nu}  \propto \nu^{-\alpha}$  with spectral  indeces $-2.07  \le \alpha  \le 0.07$ which pass  $\pm 3 \sigma$  above/below the B-band flux (shaded area). 
 If the optical  spectrum of \rbs\ were non-thermal, such a range of  PLs would be only marginally  compatible with the spectral indeces  measured  for other  types  of  isolated  neutron stars  with non-thermal optical  emission, like the rotation-powered pulsars for which  $0 \la  \alpha \la 1$  (see Mignani et  al.  2007b, 2010a,b). However, since the actual R-band flux of \rbs\  is obviously below the derived upper limit, the PL  would most likely have a negative spectral index, which makes the spectrum rising and not declining towards the near-UV.
Thus, although we  can not rule out  the presence  of a  PL component of non-thermal nature,  we regard it as unlikely that  the optical spectrum  of RBS\,  1774 can be  described by a  single PL.  Moreover, it is not clear what the origin of such a non-thermal PL  could be.  In  Zane et  al. (2008)  we  already deemed unlikely that non-thermal  optical emission from \rbs\ can be powered by the neutron  star rotational energy loss which, if  of the order of $10^{30}$  erg  s$^{-1}$ as observed in the other XDINSs, would  imply  an  anomalously large  optical emission efficiency, $\sim 1000$ times larger  than that of rotation-powered  pulsars  (Zharikov  et  al.  2006).  Alternatively, on the basis of a detection in one band only and  lacking a deep upper limit in the R band, we suggested that non-thermal  optical emission could  be powered  by the  presumably large neutron star magnetic field ($\approx 10^{14}$~G; Zane et al.  2005), as it has been proposed for the magnetars, whose optical/IR emission efficiency is also a factor of $\approx 1000$ larger than  that of the radio pulsars (Mignani et al. 2007a).
However, in the few cases where  multi-band photometry is available, magnetar optical spectra are consistent with PLs with spectral index $\alpha \ga 0$, again very much different from our  limit for \rbs.

\subsection{Thermal emission from the neutron star surface}

The R-band upper limit is consistent with a R-J spectrum  which passes  through the  B-band flux.   A  linear function connecting the  \vlt\ B-band  point and the  R-band upper limit  has a slope $\sim -1.26$,  and the R-band upper limit  lies only $\sim 25\%$ above the  extrapolation of the  R-J spectrum that passes  through the  central  value  of  the  measured  B-band  flux.   This  is  well compatible with  both the B  and R-band fluxes  being on the  same R-J spectrum, conceivably emitted  by a region of  the star surface which is colder than that  responsible for the X-ray  emission.

\begin{figure*}
\begin{center}
\includegraphics[height=7.5cm]{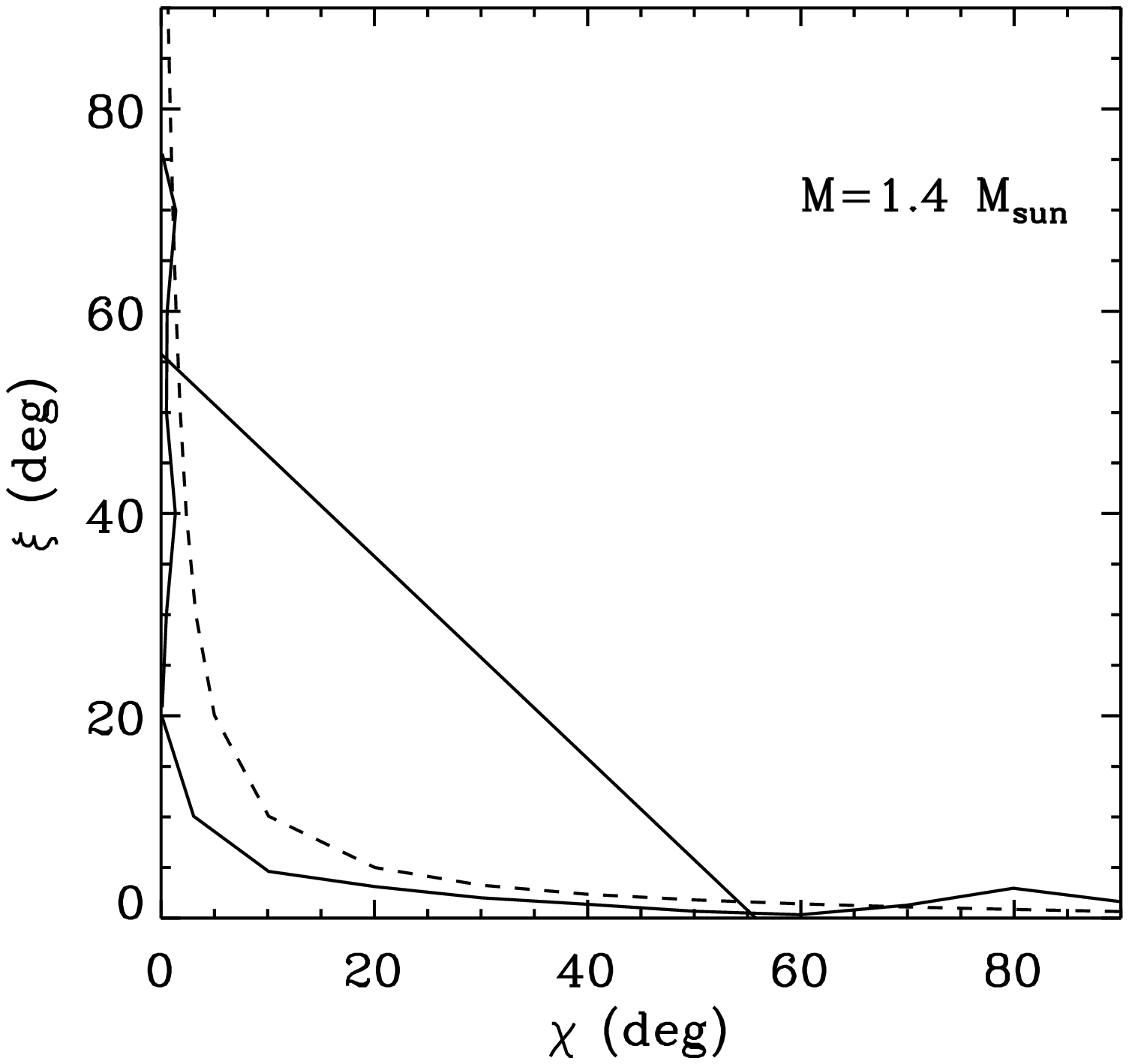}
\includegraphics[height=7.5cm]{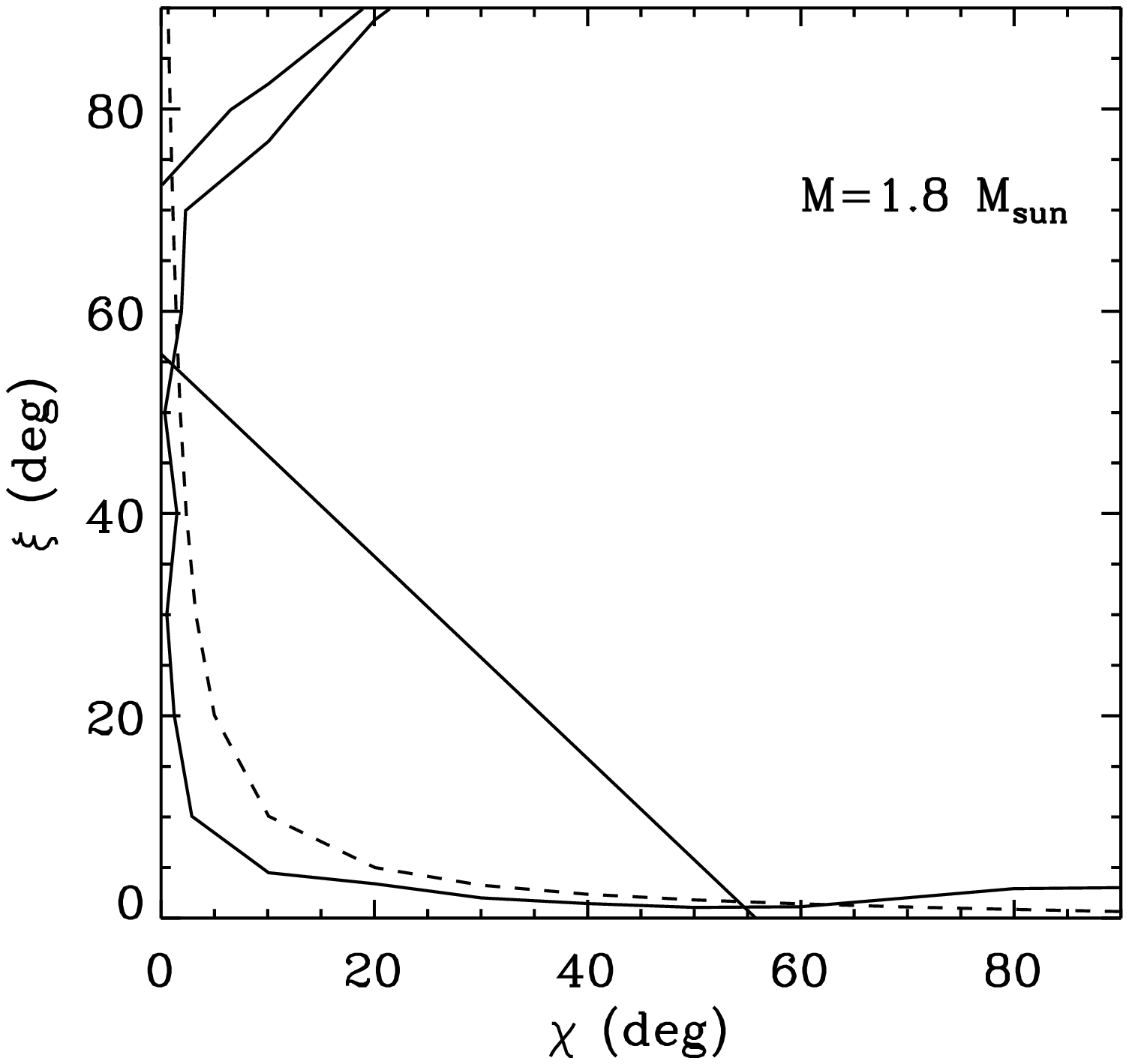}
  \caption{Left: the contour of constant PF=0.04 (solid line) for the case  $r_O=15$ km, $d=150$~pc, $T_O=11$ eV, $r_X=1$ km, and $M=1.4\, M_\odot$.  The dashed line shows the analytical result for point-like caps (see
text for details). Pulse profiles are of class I according to the classification of Beloborodov (2002), i.e. they do not
exhibit flat intervals, only when the contour is below the straight line. Right: same for $M=1.8\, M_\odot$.
 }
\end{center}
\end{figure*}

As in Zane et al. (2008), we consider a picture in  which the optical emission originates from a colder fraction of  the neutron star surface, which  emits a blackbody spectrum at a temperature $T_O$ (Braje \& Romani 2002; Pons et al. 2002; Kaplan et al. 2003; Tr\"umper et al. 2004).
 In the R-J tail,  the ratio between the  optical and X-ray  fluxes scales as $\approx r_O^2  T_O/r_X^2 T_X \equiv f  $, where $r_O$ is  the size of the optically emitting region (which, of course, cannot exceed the value
of the neutron star radius), while  $r_X$ and $T_X$ are the radius and temperature  of the  X-ray  emitting region,  as  inferred from  X-ray spectroscopy,  $T_X=104$~eV,   $r_X  =  2  (d/300   \,  {\rm  pc})$~km.
Since the \rbs\ distance is unknown,
we iterated our computations over several values of $d$ and  $r_O$ and
derived the value  of $T_O$ that corresponds  to a given value  of the optical excess, $f$ (as computed from  the B-band flux). Since no contribution from  such  a  cold  component  is observed  in  the  0.1-1~keV  \xmm\ spectrum\footnote{A soft thermal component to the X-ray spectrum could be compatible with the \xmm\ RGS data which show an emission excess at low energies (Schwope et al. 2009)}, it must be $R \ll 1$, where

\begin{equation}\nonumber 
R = \left (\frac{r_O}{r_X} \right)^2
 \left (\frac{T_O}{T_X} \right)^4
 \frac{ \int_{0.1/T_O}^{1/T_O} t^3/[\exp(t)-1] dt}{
 \int_{0.1/T_X}^{1/T_X} t^3/[\exp(t)-1] dt} \, .  
\nonumber 
\end{equation}

Furthermore, the extrapolation of  the cold blackbody component in the R band  must be compatible  with our new  \vlt\ upper limit.   We have performed the calculation  for a large set of  parameters, varying the
source  distance between  $100$ and  $500$~pc, the  B-band  excess $f$ within the $3 \sigma$  limits ($13.3\leq f\leq 35.5$), and considering $r_O=10$, $15$, and $20$~km\footnote{Neutron star radii of $\ga 20$~km are  unlikely for most equations of state (e.g., Lattimer \& Prakash 2004).}.  We found  that the data are quite constraining: a radius $r_O=15$~km is compatible with  the considered range of $f$
only for $d\leq 200$~pc (which implies $r_X  \leq 1.3$~km)  and  $T_O \leq  20$~eV. Assuming  a smaller radius  ($r_0=10$~km), the observed excess  is only compatible with    even  smaller  distances ($\sim  100$~pc,  corresponding  to $r_X=0.6$~km) and $T_0 \leq 10$~eV.  On the other hand, assuming a larger radius  ($r_O=20$~km) the observed excess is compatible with distances up to 300 pc
(corresponding  to $r_X \leq 2$~km) and $T_0 \leq 15$~eV.
 As an example, we plotted in Fig.2 a blackbody with $T_O = 11$~eV, corresponding to $r_O=15$~km and $d=150$~pc. 

Next, we verified if the inferred size of the hotter caps is  compatible with the observed pulsed fraction (PF)

\begin{equation}\nonumber 
{\rm PF}=\frac{2(F_{max}-F_{min})}{(F_{max}+F_{min})}\sim 0.04\
\nonumber 
\end{equation}

\noindent
where $F_{max}$ ($F_{min}$) are the maximum (minimum) value of the flux along the pulse
The  PF  was  computed numerically  by  means  of  an IDL  script  for different allowed combinations of $r_O$,  $r_X$ and $T_O$ (that is for which  the condition  $R\ll 1$  was met),  and assuming  in  all cases
$T_X=104$~eV.  We also  used two  values of  the star  mass, $M=1.4,\, 1.8\,  M_\odot$ to  check how  sensitive  our results  are to  general relativistic  effects  (which depend  on  the  compactness $M/R$;  e.g Beloborodov 2002). Since  it is always  $r_X\la 0.1 r_O$, we  took the latter to coincide with the star radius.
Results are shown  in Fig.3, where the curve of  constant PF (0.04) is plotted as a function of  the two geometrical angles, $\xi$, the angle between the neutron star spin  and magnetic axes, and $\chi$,  the angle between the  spin  axis  and  the  line-of-sight (LOS), 
 for the  
  case reported in Fig. 2.
 Overplotted to  the numerical  contour  is the  analytical  curve  obtained following  the method presented by Beloborodov (2002), and which is strictly valid only for point-like  caps.  The  present  analyses confirms  previous  findings (Zane et al. 2008; Schwope et al. 2009). One possibility is that  either the spin axis is nearly aligned with the magnetic axis, hence the star is a nearly aligned rotator,  or the spin axis is nearly aligned with the LOS. The other possibility is that the spin axis is slightly misaligned with respect to both the magnetic axis and the LOS by $5-10^{\circ}$.  
We note   that  although   the  numerical   and  analytical   curves  are qualitatively similar there  are quantitative differences. The allowed
range  in  the   two  angles  is  somehow  smaller   in  the  complete calculation, as it is expected  when the finite (albeit small) size of the caps  is accounted for. This also may explain the somehow different conclusions reached by Schwope et al. (2009), who used a treatment valid only for point-like caps.  The small opening  of the X-ray emitting regions 
does not constrain  the geometry too much, and several combinations of the angles $\xi$ and $\chi$ are possible.

\section{Summary and conclusions}

We performed deep optical observations of RBS\, 1774 in the R band with \fors\ at the \vlt\  to  obtain the first characterisation of  its optical spectrum.   We did not detect RBS\, 1774 candidate counterpart at its expected position
down to a $3 \sigma$ limiting magnitude of $R\sim 27$. The constraint on the colour of the candidate counterpart, $(B-R)\la0.6$, rules out  a foreground object, positionally coincident with the X-ray source. We re-analysed the \xmm\ data of \rbs\ using new calibrations and response files (see Sect. 3.2). We found that the optical  excess of  the  \vlt\ B-band  flux  (Zane et  al. 2008) with respect to the best-fit X-ray spectrum extrapolation is now $24.4 \pm 11.1$  ($3 \sigma$ confidence level). This value is still incompatible with rotation-powered non-thermal emission from \rbs, unless its optical emission efficiency  is $\sim 1000$ times larger  than that of rotation-powered  pulsars  (Zharikov  et  al.  2006).  Moreover, our R-band upper limit would most likely imply a PL spectral index $\alpha  < 0$, while rotation-powered pulsars have usually  $0\leq \alpha \leq 1$ (Mignani et al. 2007b; 2010a,b). 
On the other hand, explaining the optical excess in terms of pure thermal emission from the neutron star surface would require both a small distance
and rather stringent limits on the inclination of the LOS and the magnetic axis with respect to the neutron star spin axis.
For instance, a radius $r_O \leq 15$~km implies a neutron star  distance $d\leq 200$~pc,
which implies  $T_O \leq  20$~eV and $r_X  \leq 1.3$~km.  Such a small X-ray  emitting area is compatible with  the $\sim  0.04$  X-ray  pulsed  fraction if either
  the star spin axis is nearly aligned with the magnetic axis  or with  the LOS, or it is slightly misaligned with respect to both the magnetic axis and the LOS by $5-10^{\circ}$.
Observations in  the near-UV would  obviously be crucial to constrain the slope of the \rbs\ optical spectrum and to better constrain $T_O$  and $r_O$. With the {\em GALEX} fluxes falling well above all possible R-J spectra compatible with the \vlt\ flux measurements, only the \hstn\ (\hst) can provide the required near-UV sensitivity. Future optical observations, both from the ground and from space, will also be important to measure the source proper motion, to independently confirm the optical identification and to infer indirect constraints on the distance.

\begin{acknowledgements}
RPM thanks the ESO User Support Department and the Paranal Science Operation team for support in the scheduling and execution of the \vlt\ observations.

\end{acknowledgements}

\end{document}